\documentclass[twocolumn,showpacs,preprintnumbers,amsmath,amssymb]{revtex4}
\usepackage{dcolumn}
\usepackage{times}
\usepackage{graphicx}
\usepackage{amssymb}
\usepackage{amsmath}

\usepackage{multirow}
\newcommand{\bm}[1]{\mbox{\boldmath${#1}$}}

\newcommand{\lw}[1]{\smash{\lower1.5ex\hbox{#1}}}
\newcommand{\ds}[1]{\displaystyle{#1}}
\newcommand{\beq}{\begin{eqnarray}}
\newcommand{\eeq}{\end{eqnarray}}
\newcommand{\nn}{\nonumber}
\newcommand{\bra}{\langle}
\newcommand{\ket}{\rangle}
\newcommand{\del}{\partial}

\newcommand{\half}{\frac{1}{2}}
\newcommand{\thalf}{\frac{3}{2}}

\newcommand{\inner}[2]{{\bm #1}\cdot {\bm #2}}

\def\thline{\noalign{\hrule height 1pt}}
\begin{document}
\title{Origin of Generalized Adler-Weisberger Sum Rule for the Nucleon and $\Delta(1232)$}
\author{Keitaro Nagata}

\affiliation{
Department of Physics, Chung-Yuan Christian University, Chung-Li 320, Taiwan}

\date{\today}

\begin{abstract}
We derive a relation for the axial-coupling constants of the nucleon and 
spin-3/2 baryons including the $\Delta(1232)$ with the use of  an $SU(2)_R\times SU(2)_L$
invariant Lagrangian. Therein the nucleon belongs to 
the fundamental representation of the chiral group, while the spin-3/2 baryons 
belong to higher-dimensional representations $(1,\half)\oplus (\half,1)$. 
The resulting relation reproduces the one derived from the generalized Adler-Weisberger
(GAW) sum rule for the forward $\pi N$ scattering.
We clarify the origin of the GAW sum rule for the nucleon and $\Delta$ by taking into account both 
 the group-theoretical aspects and the spontaneous breaking of 
chiral symmetry. It turns out that the GAW sum rule for the nucleon and 
$\Delta$ is a consequence of the existence of interactions involving 
a derivative-pion-field and spontaneous chiral symmetry breaking. This implies that 
the couplings between $N$ and $\Delta$ in the GAW sum rule vanish 
with the change of the chiral condensate towards chiral restoration.
\end{abstract}

\maketitle

\section{Introduction}

It is one of recent interests to understand in-medium modifications of hadron's 
properties towards (partial) chiral restoration~\cite{Brown:2001nh}.
Recently,  STAR  at RHIC reported  
the upward mass shift of the $\Delta(1232)$~\cite{Fachini:2004jx}.
van Hees and Rapp~\cite{vanHees:2004vt} and Bi and Rafelski~\cite{Bi:2005ca} reproduced
the upward mass shift of the $\Delta(1232)$ in theoretical frameworks.
Jido, Hatsuda and Kunihiro~\cite{Jido:1999hd} and Hosaka, Dmitrasinovic and the 
author~\cite{Nagata:2008xf} extended the linear sigma model to the $\Delta(1232)$, where the mass of the $\Delta(1232)$ becomes heavier 
as the chiral condensate being an order parameter of chiral symmetry breaking becomes smaller. 
There have been several 
studies with respect to the mass shift of the nucleon $N$, e.g. ~\cite{DeTar:1988kn,Jido:2001nt,Brown:2001nh,Shuryak:2002kd,Glozman:2007ek}, along with 
another interesting question as to whether
it has any chiral partners, and what baryons are partners if any. 
Although it is still open question, most theoretical approaches agree with  that the mass of the $N$ 
becomes smaller.
These results suggest that the mass difference between the $N$ and
$\Delta(1232)$ becomes larger towards chiral restoration. 

The $\Delta(1232)$ plays an important role in the $\pi N$ scattering 
as known as the Adler-Weisberger sum rule~\cite{Weisberger:1966ip,Adler:1968hc}.
There are several suggestions that the Adler-Weisberger sum rule implies the nucleon and 
$\Delta(1232)$ to be chiral partners. If they are chiral partners, 
the mass difference becomes smaller towards chiral restoration.
Weinberg generalized the Adler-Weisberger sum rule to the algebraic relations for the axial-vector 
matrix elements by considering the asymptotic behavior, or superconvergence 
properties,  of the $\pi N$ forward scattering~\cite{Weinberg:1969hw,Weinberg:1990xn}. 
Assuming that the generalized Adler-Weisberger (GAW) sum rule is saturated by narrow resonances, the nucleon, $\Delta(1232)$ 
and the Roper resonance $N(1440)$,  
Weinberg derived relations of the masses and axial-coupling constants among $N$, $\Delta$ and 
the Roper~\cite{Weinberg:1969hw}. The resulting relations remarkably agree with the 
experimental values of the masses of the nucleon, $\Delta(1232)$ and 
the Roper. Recently Beane and van Kolck~\cite{Beane:2002ud} also investigated the baryon 
masses and axial-coupling constants with several assignment of the baryon's chiral multiplet, such as the sigma model, 
non-relativistic quark model, large $N_c$ and so on. Hosaka, Jido and 
Oka~\cite{Hosaka:2001ti} also investigated the axial-coupling constants of the nucleon and $\Delta(1232)$.
In these papers, the $N$ and $\Delta(1232)$ are mentioned as chiral partners. 
Up to this point, it seems that there is an inconsistency between the 
consequences of the GAW sum rule and the upward mass shift of the $\Delta(1232)$.

The main purposes of this paper are to clarify  underlying 
mechanisms of the GAW sum rule for the $N$ and $\Delta(1232)$
 and to solve the inconsistency on the mass shift of $\Delta(1232)$.
For this purpose, we derive the relation between the axial-coupling constants 
between the nucleon $N$ and $\Delta(1232)$ 
by using the $SU(2)_R\times SU(2)_L$ invariant Lagrangian where $N$ and $\Delta(1232)$ belong to 
different chiral multiplet.
Besides $N$, $\Delta(1232)$, we include other three spin-$\thalf$ baryons, in order to 
maintain chiral invariance. We consider $\Delta(1232)$, $\Delta(1700)$, $N(1520)$ and $N(1720)$.
This framework is refereed to 
as the quartet scheme~\cite{Hara:1965aaa,Jido:1999hd,Nagata:2008xf}. We assume $N$
to be the fundamental chiral representation $(\half,0)\oplus (0,\half)$, where $(I_R, I_L)$ is 
a representation of the (full) chiral group, and $I_R[I_L]$ are isospins for $SU(2)_R[SU(2)_L]$.
We also assume that the four kinds of spin-$\thalf$ baryons belong to a higher-dimensional 
representation $(1,\half)\oplus(\half,1)$. Then, the 
$N$ and $\Delta(1232)$ are not chiral partners in a sense that they belong to 
the different representation. 
Using this framework, we derive the relation for the axial-coupling constants of the 
baryons, assuming the existence of interaction between $N$ and $\Delta(1232)$ involving a derivative-
pion-field and of the spontaneous breaking of chiral symmetry. We find the $f_\pi$-dependence of the 
GAW sum rule for the $N$ and $\Delta(1232)$ and show that the $N$-$\Delta$ couplings 
in the GAW sum rule vanish when chiral symmetry is restored.
This result solves the inconsistency on the mass shift of $\Delta(1232)$. 

This paper is organized as follows. In the next section, we explain the framework: the 
chiral transformations of the baryons and the $SU(2)_R\times SU(2)_L$ Lagrangian. 
The chiral multiplet of the baryons are definitely determined by the chiral transformation 
properties. Then, we derive the charge-field commutation relations. 
In addition to an usual type of the charge-field commutation relations, we derive 
another types of the commutation relations causing $N$-$\Delta$ mixing. 
We show that such terms are originated from $N$-$\Delta$ interactions involving 
a derivative-pion-field and the SBCS. 
We discuss the difference between the two types of the charge-field commutation 
relations with the emphasis on the group-theoretical aspects. 
In Sec. 3,  we derive the relation among the axial-vector coupling constants of 
the baryons using the GAW sum rule. 
The final section is devoted to a summary.
%
\section{Framework}
First, we recapitulate the chiral $SU(2)_R\times SU(2)_L$ invariant Lagrangian~\cite{Nagata:2008xf}. 
Besides the nucleon $N$, we consider the four kinds of spin-3/2 baryons; 
two of them are denoted by $N_{1,2}^\mu$ and the other two 
are denoted by $\Delta^{\mu a}_{1,2}$, where the Greek and Latin indices are for Lorentz and 
isospin space.  Here $N^{\mu}_{1,2}$ and $\Delta^{\mu a}_{1,2}$ 
are basis states having the quantum numbers $I(J)^P = \half(\thalf)^+$ and $\thalf(\thalf)^+$, respectively. 
We assume that the nucleon solely belongs to the fundamental representation 
$(\half,0)\oplus (0, \half)$, and that $\Delta(1232)$ belongs to a chiral 
representation $(\half,1)\oplus (1,\half)$ together with other three spin-3/2 
baryons. 
The $SU(2)_A$ transformations are given by 
\begin{subequations}
\begin{align}
\delta_5^{\vec{a}} N =&\half i\gamma_5 \inner{\tau}{a} N,\\
\delta_5^{\vec{a}} N_{\rm n}^\mu =& \frac{\eta}{2} \left( \frac53 i \inner{a}{\tau} \gamma_5
N_{\rm n}^\mu + \frac{4}{\sqrt{3}} i \gamma_5 \inner{a}{\Delta_{\rm n}^\mu}\right), \;\; ({\rm n}=1,2),
\label{eq:chiralnmu}\\
\delta_5^{\vec{a}} \Delta_{\rm n}^{\mu i} =& \frac{\eta}{2}\left(\frac{4}{\sqrt{3}} i
\gamma_5 a^j P^{ij}_{3/2} N_{\rm n}^\mu
- \frac23 i \tau^i \gamma_5 \inner{a}{\Delta_{\rm n}^\mu}\right.\nn \\ 
 &\left.+ i\inner{a}{\tau}\gamma_5\Delta_{\rm n}^{\mu i}\right).
\label{eq:chiraldelmu}
\end{align}%
\label{eq:su2a1}%
\end{subequations}%
Here $\tau^i$ is the Pauli-matrices for isospin, $P^{ij}_{3/2}= \delta^{ij}-\tau^i\tau^j/3$ is the isospin-3/2 projection 
operator, and $a^i$ are the infinitesimal parameters for the $SU(2)_A$ transformation. 
We introduce $\eta=+1 $ for $(N_1^\mu, \Delta_1^{\mu i})$ and $\eta=-1$ for $(N_2^\mu, 
\Delta_2^{\mu i})$.
We omit the $SU(2)_V$ transformation, because all the baryon fields belong to 
the irreducible representations of isospin. 
The chiral transformation of the nucleon is closed within $N$ itself, and 
does not include any other baryon-field. On the other hand, a pair of $I=\half$ and 
$\thalf$ states, $(N_1^\mu, \Delta_1^{\mu a})$, forms an irreducible representation. Hence their 
co-existence is necessary for chiral invariance. The other pair $(N_2^\mu, \Delta_2^{\mu a})$ 
also has the same chiral transformation laws, but has the opposite sign to $(N_1^\mu, \Delta_1^{\mu a})$. 

These basis states $N$, $N_{1,2}^\mu$ and $\Delta_{1,2}^{\mu i}$ are assigned with the observed baryons by considering SBCS.
Note that the relative difference in sign $\eta=\pm 1$ is important to obtain the masses of the 
observed states. 
For the nucleon $N$, we employ the Gell-Mann-Levy type Lagrangian. 
For the spin-3/2 sector, chiral invariant interaction terms are given by 
\begin{subequations}
\begin{align}
{\cal L}_{\pi B B}^{(1)}&=  g_1\left( \bar{\Delta}_{1\mu}^i U_5 \Delta_1^{\mu i}
-\frac{3}{4}\bar{N}_{1\mu} U_5 N_1^\mu \right. \nn \\
&+\left.\frac{1}{12} \bar{N}_{1\mu} \tau^i U_5 \tau^i N_1^\mu
+\frac{\sqrt{3}}{6}\bar{N}_{1\mu} \tau^i U_5 \Delta_1^{\mu i} \right),
\label{eq:chiralint1}\\
{\cal L}_{\pi B B}^{(2)} &=  g_2\left( \bar{\Delta}_{2\mu}^i
U_5^\dagger \Delta_2^{\mu i}
-\frac{3}{4}\bar{N}_{2\mu} U_5^\dagger N_2^\mu \right. \nn \\
&+ \left.\frac{1}{12} \bar{N}_{2\mu} \tau^i U_5^\dagger \tau^i
N_2^\mu +\frac{\sqrt{3}}{6}\bar{N}_{2\mu} \tau^i U_5^\dagger
\Delta_2^{\mu i} \right),
\label{eq:chiralint2}\\
{\cal L}_{BB} &= - m_0 \left(\bar{\Delta}_{1 \mu }^i \Delta_2^{\mu
i} + \bar{N}_{1\mu} N_2^\mu \right),
\label{eq:chiralint3}
\end{align}%
\label{Dec1212eq2}%
\end{subequations}%
where $U_5=\sigma+i\gamma_5 \vec{\tau}\cdot\vec{\pi}$. These terms 
generate the masses of the baryons. 
Following Ref.~\cite{Jido:1999hd,Nagata:2008xf}, we consider 
the observed baryons $N(1520), N(1720)$ and $\Delta(1700)$ as well as $\Delta(1232)$.

Having the transformation laws Eqs.~(\ref{eq:su2a1}), we derive the 
Neother current and charge-field commutation relations in a straightforward
manner. 
The axial-vector current $A^a_\mu$ is defined by
\begin{align}
A^a_\mu = -i \frac{\del {\cal L}}{\del(\del^\mu B_n)} t^{a}_{nm} B_m,
\label{Sep1308eq2}
\end{align}
where $B_n$ denotes baryon fields $N$, $\Delta_{1,2}^{\mu i}$ and $N^{\mu}_{1,2}$, and 
the matrices $t_{nm}^i$ is defined through the axial-vector transformation
$\delta^{\vec{a}}_5 B_n = i a^i t^{i}_{nm} B_m$.
We obtain the axial-vector current;
\begin{subequations}
\begin{align}
A^{\mu a}_{(1)} &= \half \bar{N}\gamma^\mu \gamma_5 \tau^a N,
\label{sep2408eq1} \\
A^{\mu a}_{(2)} &= \frac{\eta}{2} \left[\frac{5}{3} \bar{N}_{{\rm n}\beta} \gamma^\mu \gamma_5 \tau^a N^\beta_{\rm n}
  +\bar{\Delta}_{{\rm n}\beta}^b \gamma^\mu\gamma^5 (\tau^a) \Delta^{\beta b}_{\rm n} \right. \nn\\
&+\left. \frac{4}{\sqrt{3}} \left( \bar{N}_{{\rm n}\beta} \gamma^\mu \gamma_5 \Delta^{\beta a}_{\rm n} + (h.c.)\right) 
\right].
\label{sep2408eq2}
\end{align}%
\label{Sep2008eq1}%
\end{subequations}%
Note that $A^{\mu a}_{(1)}$ and $A^{\mu a}_{(2)}$ come from the kinetic terms and 
therefore that they do not contain any free parameters. Now the charge-field commutation 
relations are given, with $Q_A^a =\int d^3x A^{a}_{\mu=0} (x)$, by 
\begin{subequations}
\begin{align}
[Q_A^a, N(x)] &=  \half \tau^a \gamma_5 N(x), 
\label{Dec0108eq1} \\
[Q_A^a,\Delta^{\mu b}_n(x)] &= \eta\left[ \half \gamma_5 \tau^a \Delta^{\mu b}_n +\frac{2}{\sqrt{3}} \gamma_5 \delta^{ab} N^\mu_n\right],
\label{Dec0108eq2} \\
[Q_A^a, N^\mu_n (x)] &=\eta \left[ \frac{5}{6} \gamma_5 \tau^a N^\mu_n + \frac{2}{\sqrt{3}} \gamma_5 \Delta_n^{\mu a }\right]
\label{Dec0108eq3} 
\end{align}%
\label{Dec0108eq4}%
\end{subequations}%
These transformation properties determine the axial-charges of the baryons~\cite{Nagata:2007di}.
Clearly, $N_1^\mu$ and $\Delta_1^{\mu a}$ are a member of the identical chiral multiplet and 
chiral partners.

Next, we consider interaction terms between the nucleon and spin-3/2 baryons; 
there are three terms, 
\begin{subequations}
\begin{align}
{\cal L}_{\pi N B}^{(1)}&= \frac{g_3}{\Lambda^2}
\left[\bar{N} U_5 (\del_\mu \pi^i) \Delta^{\mu i}_1 \right. \nn \\ 
 &\left. + \frac{\sqrt{3}}{2} \bar{N} U_5 ( -\gamma_5(i\del_\mu \sigma)
 + \frac{1}{3}\del_\mu \inner{\pi}{\tau})N^\mu_1\right],
\label{eq:chiralint4}\\
{\cal L}_{\pi N B}^{(2)} &= \frac{g_4}{\Lambda^2} \left[\bar{N}
(\del_\mu U_5) ( \pi^i)\Delta^{\mu i}_1  \right. \nn \\ 
 &\left.+\frac{\sqrt{3}}{2} \bar{N} (-i\del_\mu U_5) (\gamma_5 \sigma
+\frac{i}{3} \inner{\pi}{\tau})N^\mu_1\right],
\label{eq:chiralint5}\\
{\cal L}_{\pi N B}^{(3)} &=   \frac{g_5}{\Lambda}
\left[\bar{N}(\del^\mu \pi^i) \Delta_{2\mu }^i  - \frac{\sqrt{3}}{2}\bar{N} (-i\del^\mu)(\gamma_5\sigma-\frac{1}{3}
i \inner{\tau}{\pi})N_{2\mu} \right].
\label{eq:chiralint6}
\end{align}%
\label{Dec1212eq1}%
\end{subequations}%
Note that each ${\cal L}_{\pi N B}^{(1)}$, ${\cal L}_{\pi N B}^{(2)}$ and ${\cal L}_{\pi N B}^{(3)}$
is chiral invariant. Here we introduce a parameter $\Lambda$ with dimension [mass] 
to keep the coupling constants $g_3$, $g_4$ and $g_5$ dimensionless. In fact, $\Lambda$ 
can be absorbed into the definitions of $g_{3,4,5}$ and is irrelevant of the following discussions.
For the derivation of these interaction terms, see Ref.~\cite{Nagata:2008xf}.

These interaction terms Eqs.~(\ref{Dec1212eq1}) contribute to the axial-vector current, because
they contain  a derivative-pion-field $(\del^\mu \pi^i)$. 
In the linear realization of chiral symmetry, the chiral transformation of the pion field is given by its 
chiral partner, i.e., $\sigma$; $\delta_5^{\vec{a}} \pi^i = - a^i \sigma$. 
Therefore, their contributions are given by
\begin{align}
A^{\mu a}_{(3)}&= 
\frac{g_3}{\Lambda^2} \sigma^2 \bar{N} \Delta^{\mu a}_1 +\frac{\sigma^2}{2\Lambda^2} \left( g_3 \frac{\sqrt{3}}{3} + \sqrt{3} g_4 \right) 
\bar{N} \tau^a N_1^\mu  \nn \\
+&\frac{g_5 }{\Lambda} \sigma\left( \bar{N} \Delta_{2 \mu} ^a + \frac{\sqrt{3}}{6}\bar{N} \tau^a N_2^\mu\right).
\label{Dec2008eq1}
\end{align}%
Here we do not take into account four-point coupling terms containing pion-fields such as $N\Delta\sigma\pi$ terms, 
because they are irrelevant of the following discussions.
Note that $A^{\mu a}_{(3)}$ depends on the interaction strengths $g_3, g_4, g_5$, which are
determined from physical conditions but by chiral symmetry.  
This is in contrast to $A^{\mu a}_{(1)}$ and $A^{\mu a}_{(2)}$, which do not 
contain any free parameters.
Note also that $A^{\mu a}_{(3)}$ contains three- and four-point couplings. 
This is also different from that $A^{\mu a}_{(1),(2)}$ derived from kinetic terms are two-point 
couplings. The two-point coupling terms contribute to 
the charge-field commutation relations as one-particle states, while 
the three-point couplings contribute to as multi-particle states. 
Nevertheless, $A_{(3)}^{\mu a}$ becomes two-point couplings if chiral symmetry 
is spontaneously broken.

In order to see how the interaction terms contribute to the axial-vector current and 
to the charge-field commutation relations, we first consider a simple case including only $\Delta_1$.
With SBCS the sigma field has vacuum expectation values $\bra \sigma\ket \to f_\pi$.
Then, Eq.~(\ref{Dec2008eq1}) is reduced to
\begin{align}
A^{\mu a}_{(3)} = \frac{g_3 f_\pi^2}{\Lambda^2}  \bar{N} \Delta_1^{\mu a},
\label{Dec0808eq1}
\end{align}
which is two-point coupling causing the transition between the 
nucleon and $\Delta_1$. Note that 
Eq.~(\ref{Dec0808eq1}) involves chiral order parameter $f_\pi$. This implies that 
the nucleon-$\Delta$ axial-vector coupling results from the spontaneous breaking of 
chiral symmetry. 
 Including full terms in Eq.~(\ref{Dec2008eq1}), we obtain
 \begin{subequations}
\begin{align}
[Q_A^a, N] & = \frac{f_\pi}{\Lambda} g^{\mu 0} \gamma^0 \left( g_{\pi N\Delta^+} \Delta_{+\mu}^a + g_{\pi N \Delta^-} \gamma_5 
\Delta_{-\mu }^a \right)\nn \\
&+ \frac{f_\pi}{\Lambda} \tau^a g^{\mu 0} \gamma^0 ( g_{\pi N N^{*+}} N_{+\mu} + g_{\pi NN^{*-}} \gamma_5 N_{-\mu} )\\
[Q_A^a, N_+^\mu] &= \frac{f_\pi}{\Lambda} g_{\pi NN^{*+}}  g^{\mu 0}\gamma^0 \tau^a N, \\
[Q_A^a, N_-^\mu] &= \frac{f_\pi}{\Lambda} g_{\pi NN^{*-}} \gamma_5  g^{\mu 0}\gamma^0 \tau^a N, \\
[Q_A^a, \Delta_+^{\mu b}] & = \frac{f_\pi}{\Lambda} g_{\pi N \Delta^+} g^{\mu 0} \gamma^0 \delta^{ab} N, \\
[Q_A^a, \Delta_-^{\mu b}] & = \frac{f_\pi}{\Lambda} g_{\pi N \Delta^-} \gamma_5 g^{\mu 0} \gamma^0 \delta^{ab} N.
\end{align}%
\label{Dec1212eq3}%
\end{subequations}%
Here we redefine the baryon fields as mass eigenstates, where the masses 
are generated by Eqs.~(\ref{Dec1212eq2}). The mass eigenstates are given by 
\begin{subequations}
for $\Delta$s
\begin{align}
\Delta_{+}^{\mu i} &= \frac{1}{\sqrt{2}} (\Delta_1^{\mu i} + \Delta_2^{\mu
i}),\\ 
\Delta_{-}^{\mu i} &= \frac{1}{\sqrt{2}}  \gamma_5 (-\Delta_1^{\mu
i}+\Delta_2^{\mu i}), 
\end{align}
and for the $N^*$s,
\begin{align}
N_{-}^{\mu} &=  \frac{1}{\sqrt{2}}  \gamma_5 (-N_1^\mu + N_2^\mu),\\
N_{+}^{\mu} &= \frac{1}{\sqrt{2}}  (N_1^\mu + N_2^\mu),
\end{align}%
\end{subequations}%
where the subscripts $\pm$ denote the parity. 
The four $\pi N$ coupling constants are given by $g_{3,4,5}$, 
\begin{subequations}
\begin{align}
g_{\pi N \Delta^\pm} &=\frac{1}{\sqrt{2}\Lambda}(g_5\Lambda \pm g_3 f_\pi), \\
g_{\pi N N^{*\pm}} &=\frac{\sqrt{6}}{12\Lambda}\left(g_5\Lambda \pm  (g_3 + 3g_4) f_\pi
\right).
\end{align}%
\end{subequations}%
All terms in  Eqs.~(\ref{Dec1212eq3}) depend both on $f_\pi$ and interaction strengths due to 
its derivation from the interaction terms. 
Hence the origin of Eqs.~(\ref{Dec1212eq3}) is in the existence of the 
interaction terms with a derivative-pion-field and the SBCS.
Then, all terms in Eqs.~(\ref{Dec1212eq3}) vanish 
when chiral symmetry is restored $\bra \sigma \ket = f_\pi \to 0$.
Despite of the absence of the mixing between the nucleon and $\Delta$ under 
$SU(2)_A$ transformations, we derived the two-point coupling 
terms between the nucleon-spin-3/2 baryon in the axial-vector current, such as $N$-$\Delta$ couplings,  and the mixing terms in the 
charge-field commutation relations.
Equations (\ref{Dec1212eq3}) tell us several important consequences;
\begin{enumerate}
\item The $N$-$\Delta$ mixings in the axial-vector current results from the pion-derivative interactions and SBCS.
\item The strengths of the mixings are  determined by the physical consequences, but by a chiral representation. 
\item The number of states joined this mixings is not limited by a multiplicity of the chiral representation.
\item The mixings vanish when chiral symmetry is restored. 
\item Hence the degeneracy between $N$ and $\Delta$ 
is not required.
\end{enumerate}
These points are different from the well known consequences of 
partners in group-theoretical sense such as the proton and 
neutron, i.e., an isodoublet of $SU(2)_V$. Partners defined in the group-theoretical sense belong to an identical 
multiplet and the number of the members belonging to the multiplet is given by the multiplicity of the multiplet. 
The conserved charges are uniquely determined corresponding to the irreducible representation.
In addition, the partners degenerate when symmetry is restored. 
Contrarily, there are dynamical consequences behind the $N$-$\Delta$ 
mixing in the axial-vector current.  Although some of these points are already 
known, it is new to clarify that the origin of the coupling is in  
the existence of the pion-derivative interactions and SBCS.

In the derivation of Eqs.~(\ref{Dec1212eq3}), we use the Lagrangian (\ref{Dec1212eq2}). 
However, the consequences obtained above do not depend on the details of 
the interaction terms. Let us consider an interaction term
\begin{align}
{\cal L}_{\rm int} = g\bar{\phi}^a \Gamma^{\mu  i}_{ab} \psi^b (\del_\mu \pi^i), 
\end{align}
where $\phi$ and $\psi$ can be {\it any} baryon fields and  indices $a$ and $b$ label
quantum numbers of the baryon fields such as spin and isospin. 
A matrix $\Gamma^{\mu i}_{ab} $ is a function of the Dirac- and Pauli-matrices and 
chiral order parameter $\bra \sigma\ket$. Then, this term contributes to the 
axial-vector current as 
\begin{align}
A^{\mu i} = \frac{\del {\cal L}_{\rm int} }{\del (\del_\mu \pi^i ) } \sigma \overset{{\rm SBCS}}{\to}
f_\pi g \bar{\phi}^a \Gamma^{\mu}_{ab} \psi^b.
\end{align}
Again, this term gives the mixing between $\phi$ and $\psi$ in the charge-field commutation 
relations. Note that we do not need to specify chiral representation of $\phi$ and $
\psi$ in order to derive the mixing. The axial-vector current originating from the interaction terms 
necessarily involves VEVs of the sigma field $\bra \sigma \ket \neq 0$ 
due to the pion's axial-vector transformation.
We can include any baryon fields,  assuming the existence of an interaction accompanying a derivative-pion-field.

\section{$N$-$\Delta$ relation for the generalized Adler-Weisberger Sum rule}

Now, we consider an application with the use of the charge algebra
\begin{align}
[Q_A^a, Q_A^b] = i\epsilon^{abc} T^c,
\label{Dec0608eq1}
\end{align}
where $T^a$ are $SU(2)_V$ generators and $Q_A^a$ are $SU(2)_A$ generators.
We take a matrix element of Eq.~(\ref{Dec0608eq1}) with the initial and final nucleon states carrying 
the momentum $\vec{p}$ and $\vec{p}^{\;\prime}$;
\begin{align}
\bra N(\vec{p}^{\;\prime}) |[Q_A^a, Q_A^b] | N(\vec{p}) \ket =\bra N(\vec{p}^{\;\prime}) | (i\epsilon^{abc} T^c)| N(\vec{p}) \ket.
\end{align}
Now we define the axial-vector coupling matrix between the initial baryon with momentum $\vec{p}$ and 
the final baryon with $\vec{p}^{\; \prime}$ by 
\begin{align}
X^a = \bra B(\vec{p}^{\;\prime})| Q^a_A |B(\vec{p})\ket.
\end{align}
Here we ignore the pion pole term $(f_\pi \del_\mu \pi^a)$ from the axial-vector current 
in the definition of the axial-vector coupling matrix.
Then, we can rewrite Eq.~(\ref{Dec0608eq1}) as 
\begin{align}
[X^a, X^b] = i\epsilon^{abc} T^c,
\label{sep2408eq6}
\end{align}
which was derived by considering the superconvergence property of the forward 
$\pi$-$N$ scattering~\cite{Weinberg:1969hw,Hosaka:2001ti,Beane:2002ud}. 
In the literature~\cite{Weinberg:1969hw,Weinber:1990xn,Beane:2002ud}, Eq.~(\ref{sep2408eq6}) is refereed to as the generalized 
Adler-Weisberger sum rule.

Following Ref.~\cite{Weinberg:1969hw,Weinberg:1990xn,Beane:2002ud}, we assume that the generalized Adler-Weisberger 
sum rule is saturated only by narrow resonances. In the present work, we 
consider the the nucleon and four spin-3/2 baryons as intermediate states, and do not 
include other resonances; 
\begin{align}
1= |N \ket \bra N | +\sum_R | R \ket \bra R |,
\end{align}
where $R= \Delta^{\pm}, N^{*\pm}$. 
Taking the infinite momentum frame $\vec{p}^{\;\prime} \sim \vec{p} \to \infty $ and using 
the Gordon identity, we obtain
\begin{align}
g_A^2- g_V = \frac{4}{9}\sum_{R=\Delta^\pm} {\cal C}_{RN} -\frac{4}{3}\sum_{R=N^{*\pm}} {\cal C}_{RN}.
\label{eq:Dec0608eq2}%
\end{align}%
Here we introduce $g_V$ and $g_A$  for the nucleon vector and 
axial-vector coupling constant in order to separate the contributions from nucleon and other baryons, 
and ${\cal C}_{RN}$ is the contributions from the spin-3/2 baryons $R=\Delta^{\pm}, N^{*\pm}$, 
\begin{align}
{\cal C}_{RN}= f_\pi^2 \frac{M_N}{M_{R}} \left(\frac{g_{\pi N R}}{\Lambda}\right)^2
\end{align}
In the standard Gell-Mann Levy model $g_A=1$, and the deviation from $g_A=1$ can be obtained by
the mixing of other chiral multiplets or the pion-derivative coupling~\cite{Jido:2001nt}, while 
$g_V=1$. 
\begin{table}[tbh]
\begin{center}
\caption{The contributions ${\cal C}_{RN}$ for each baryons $R=\Delta^{\pm}, N^{*\pm}$. 
In the last column, we show the sum of the four contributions, i.e., r.h.s of Eq.~(\ref{eq:Dec0608eq2}) and 
 the experimental value of l.h.s. $g_A^2-g_V$.
We showed the values of the baryon masses and $\pi N$ coupling constants obtained 
by using the linear sigma model. For details, see Ref.~\cite{Nagata:2008xf}.} 
\begin{tabular}[t]{cccc}
\thline
States $R$  & ${\cal C}_{RN} $ & $M_R$ [MeV] & $\ds{\frac{g_{\pi N R}}{\Lambda}}$ [MeV$^{-1}$]\\
\hline 
$\Delta^+$ [$\Delta(1232)$]  & 2.00  & 1320  & 18\\
$\Delta^-$ [$\Delta(1700)$]  & 0.459 & 1770 & 10 \\
$N^{*-}$ [$N(1520)$] & 0.212 & 1430 & 6.1\\
$N^{*+}$ [$N(1720)$] & 0.024 & 1660 & 2.2 \\
\hline
Sum & 0.78 (exp 0.54) & - & -\\
\thline
\end{tabular}
\label{tab:Dec2008tab1}
\end{center}
\end{table}
We show the result in Table.~\ref{tab:Dec2008tab1}. The $\Delta(1232)$ is dominant, and 
$\Delta^-$ and $N^{*-}$ also have non-negligible effects. The contribution of $N^{*+}$ 
is almost negligible. The sum of the four contributions is ${\cal C}_{RN} = 0.78$. The experimental 
values of the l.h.s. of Eq.~(\ref{eq:Dec0608eq2}) is $g_A^2-g_V=0.54$.
Considering the absence of other resonances in the present framework, this is reasonable 
agreement. Beane showed that among various resonances the Roper has the largest contribution, 
and reduces the r.h.s of Eq.~(\ref{eq:Dec0608eq2})~\cite{Beane:2002ud}. 
Hence the inclusion of the Roper and other resonances will improve the present 
result. 
Beane and van Kolck also showed that the inclusion of all the four- and three-star resonances 
in PDG quantitatively reproduces physical values of $g_A$.
Therefore we expected that there is no other state coupling to $\pi N$ 
strongly. This is consistent with a behavior about that $\pi N$ decays 
of highly excited states are small in spite of the large phase space.
If any baryons decay to $\pi N$ strongly, it would have other partners to saturate the GAW 
sum rule.
Such partners would be in the pair of isospin 1/2 and 3/2 states, as shown 
in Eq.~(\ref{Dec0608eq1}). 
Although some of these results have already know, the new results are the derivation 
of the GAW sum rule for $N$ and $\Delta$ by using the field theoretical framework with $SU(2)_R\times SU(2)_L$ 
symmetry and the findings of the $f_\pi$ dependence of the GAW sum rule.

\section{Summary}

We have derived the relation for axial-coupling constants of the nucleon and 
four types of spin-3/2 baryons, $\Delta(1232)$, $\Delta(1700)$, $N(1520)$ and $N(1720)$, 
by using the $SU(2)_R\times SU(2)_L$ 
invariant Lagrangian and the generalized Adler-Weisberger sum rule. 
We assigned the nucleon with the fundamental representation, 
and four spin-3/2 baryons with $(1,1/2)\oplus (1/2,1)$ and 
therefore, the nucleon and $\Delta(1232)$ belong to different 
chiral multiplets. We clarified that the origin of the mixings between the nucleon 
and $\Delta(1232)$ in the GAW sum rule is in the existence of the interaction terms accompanying a
derivative-pion-field, and of the spontaneous breaking.
This suggests that the $N$ and $\Delta$ are not chiral partners and 
that it is not necessary that they degenerate towards chiral restoration. 
This also suggests that the mass difference between $N$ and $\Delta(1232)$ 
can be larger towards chiral restoration in a manner consistent with the GAW sum rule.
These results partly support the upward mass shift of the $\Delta(1232)$~\cite{Fachini:2004jx,vanHees:2004vt,Bi:2005ca,Jido:1999hd,Nagata:2008xf}.

Finally, we comment on that the present paper does not exclude the possibility that 
the nucleon and $\Delta(1232)$  are chiral partners. In order to realize this possibility, 
it is necessary to find a multiplet including both the nucleon and $\Delta(1232)$.
This is a longstanding problem as to whether 
the different spin states can be a member of a multiplet~\cite{Hara:1965aaa}. 
One of the simple approaches to put $N$ and $\Delta$ into one multiplet is 
the non-relativistic quark model with $SU(2 N_F)$ spin-flavor symmetry 
or large $N_c$. On the other hand, in a relativistic framework,  we have 
shown~\cite{Nagata:2007di,Chen:2008qv} that there is no such chiral 
multiplets that contains both the nucleon and $\Delta(1232)$ at least for
three-quark levels. The direct proof  that $N$ and $\Delta$ are 
chiral partners is to find a chiral multiplet including both the 
nucleon and $\Delta(1232)$.

The author acknowledge to A. Hosaka for a valuable comment and H. Hatanaka and C. W. 
Kao for fruitful discussions.
The author is supported by National Science Council (NSC) of Republic 
of China under grants No. NSC96-2119-M-002-001.


\end{document}